\documentclass[aps]{revtex4-1}
\usepackage[utf8]{inputenc}
\usepackage{amsmath,amssymb}
\usepackage{graphicx}

\newcommand{\Bomega}{ \bar{\Omega} }

\begin{document}
\title{\textbf{Soap bubble hadronic states in a QCD-motivated
Nambu-Jona-Lasinio model}}

\author{Sergii Kutnii}
\affiliation{Bogolyubov Institute for Theoretical Physics, Kyiv, Ukraine}
\email{mnkutster@gmail.com}

\begin{abstract}
Inhomogeneous solutions of the gap equation in the mean field approach to
Nambu-Jona-Lasinio model are studied.
An approximate Ginzburg-Landau-like gap equation is obtained and the domain wall
solution is found. Binding of fermions to the domain wall is demonstrated.
Compact domain wall with bound fermions is studied and stabilisation by fermion
pressure is demonstrated which opens a possibility for existence of ``soap
bubble'' hadronic states.
\end{abstract}

\begin{titlepage}

\maketitle

\end{titlepage}

\tableofcontents

\newpage

\section{Introduction}

Building analogies between fundamental field theories and condensed matter
physics has been a long-standing trend. Usually, quantum field theory methods developed in
fundamental interactions research were applied to condensed matter problems.
However, a flow of ideas in the opposite direction exists as well. One of the
fields influenced by this flow is the physics of strong interactions at low
energies. As the perturbative QCD computation breaks down in this region, other
methods are necessary.

Savvydy \cite{Savvydy} has demonstrated instability of the QCD vacuum with
zero field strength in 1977. Since then existence of various vacuum condensates
in QCD became well established. This suggested looking for analogies with
condensed matter models where such condensation occurs, superconductivity being
the primary choice. Analogy with superconductivity gave rise to two models, one
being the dual superconductivity \cite{Ripka} which is not our topic and the
Nambu-Jona-Lasinio (NJL) model being the other.

The Nambu-Jona-Lasinio model was first proposed in 1961 \cite{NJL}, well before
QCD. Starting from the fermionic lagrangian built as an analogy to the BCS one

\begin{equation}
\mathcal{L} = i\bar{\psi}\widehat{\partial}\psi +
g\left[\left(\bar{\psi}\psi\right)^2 -
\left(\bar{\psi}\gamma_5\psi\right)^2\right],
\label{eq:NJLmodel}
\end{equation}

Nambu and Jona-Lasinio demonstrated that fermion condensation occurs here as
well.

Since then, their model has been mostly used to describe mesons (see e.g.
\cite{Volkov1}, \cite{Volkov2}). Attempts to look beyond the simple homogeneous
case studied by the model's authors were rare (one notable exception being
\cite{BasarDunne}).

However, studying inhomogeneous cases is important as a number of interesting
phenomena may occur. Consider the NJL gap equation with an Euclidean cutoff at
$p^2 = \Lambda$:
\begin{equation}
\frac{2\pi^2}{g\Lambda^2} = 1 -
\frac{M^2}{\Lambda^2}\ln\left[\frac{\Lambda^2}{M^2} + 1\right]
\end{equation}

This equation depends on $M^2$ only, which means it allows two
symmetric opposite-sign solutions.

This means that the model has two possible true vacuum states. Thus, there
exists a possibility for existence of spatial regions with different vacuum
values separated by domain walls.

What happens to a fermion in such a vacuum? To answer this question, we note
that as the condensate is obtained in the mean field approximation,
the equation for fermionic states will be

\begin{equation}
\left[i\widehat{\partial} - M(x)\right]\psi = 0.
\end{equation}

In the flat case where $M$ depends on only one spatial coordinate this is
exactly the Jackiw-Rebbi problem \cite{JackiwRebbi}. In their 1976 work Jackiw
and Rebbi have demonstrated existence of a $E=0$ fermionic
state with $1/2$ fermion number. In our case where $M(x)$
interpolates between $\pm{M_\infty}$ this state's asymptotic behaviour will be
$\exp\left(\mp{M_\infty}x\right)$ which means it will be bound to the domain
wall.

But what if the domain wall is not infinite but encloses a small bubble
of $-M$ vacuum in the $+M$ sea instead? From general considerations we can
expect that the bubble's energy will be proportional to its surface area so it
will not be stable but if there are fermions trapped at its surface their
pressure may have a stabilising effect. If such stable bubbles exist, they are
natural candidates for hadrons.

Our paper is organised as follows: in section 2 we repeat our results
first presented in \cite{Inhomogen} where an approximation of the NJL
gap equation that allows us to study inhomogeneous vacuum states was developed;
in section 3 our solution to the Jackiw-Rebbi problem with a single flat 
domain wall solution of the approximate gap equation from \cite{DomainWall}
is presented and in section 4 we study the vacuum bubble and demonstrate that
under certain conditions it gets stabilised.

\section{The mean field dynamics}

The initial quark action of our model is

\begin{equation}
S = \int{}d^4x\left[\bar{\psi}i\widehat{\partial}\psi +
\frac{1}{4G^2}\bar{\psi}\gamma_{\mu}T^a\psi\bar{\psi}\gamma^{\mu}T^a\psi\right]. \label{eq:defmodel}
\end{equation}

Here $\psi$ bears the color index $i$ as well as the spinor one and the
generators $T^a$ of the $SU(N)$ algebra are defined conventionally and obey the relations:

\begin{eqnarray}
&&T^a{}T^b = \frac{1}{2N_c}\delta^{ab} + \frac{1}{2}\left(if^{abc}T^c +
d^{abc}T^c\right); \nonumber\\
&&d^{abc}d^{abd} = \frac{N_c^2 - 4}{N_c}\delta^{ad}.
\end{eqnarray}

$N_c$ is the number of colors. We keep it arbitrary for the sake of possible generalizations.

The model (\ref{eq:defmodel}) is not the Nambu-Jona-Lasinio model defined in \cite{NJL}. 
However, its interaction is just the quark currents coupling and it's intuitive to assume 
that if tha quarks are integrated out from the system somehow, 
then the resulting quark-quark interaction will look like this in the first order at least.

We have proposed a derivation of such an action from the QCD lagrangian in \cite{Me1}, indeed. 
A similar quark interaction arises when instanton condensate is present (see \cite{InstCond}).

The path integral of the model is equal to

\begin{eqnarray}
&&Z = \int\mathcal{D}\Omega\mathcal{D}\bar{\psi}\mathcal{D}\psi\exp\left\{
-i\int{}d^4{}x\left[\bar{\psi}i\widehat{\partial}\psi +
\frac{1}{4G^2}\bar{\psi}\gamma_{\mu}T^a\psi\bar{\psi}\gamma^{\mu}T^a\psi
-\right.\right.\nonumber\\
&&\left.\left. - Sp\left[\gamma_{\mu}T^a\left(G\Omega +
\frac{1}{2G}\psi\bar{\psi}\right)\gamma^{\mu}T^a\left(G\Omega +
\frac{1}{2G}\psi\bar{\psi}\right)\right]\right]\right\} = \nonumber\\
&&= \int\mathcal{D}\Omega\mathcal{D}\bar{\psi}\mathcal{D}\psi\exp\left\{
-i\int{}d^4{}x\left[\bar{\psi}i\widehat{\partial}\psi -
\right.\right.\nonumber\\
&&\left.\left. - \bar{\psi}\gamma_{\mu}T^a\Omega\gamma^{\mu}T^a\psi -
G^2Sp\gamma_{\mu}T^a\Omega\gamma^{\mu}T^a\Omega\right]\right\}.
\label{eq:modaction}\end{eqnarray}
up to a constant factor. 
The new variable $\Omega$ should have just enough degrees of freedom to couple to $\psi\bar{\psi}$ in the action (\ref{eq:modaction}). This means that $\Omega$ can be written as follows:
\begin{eqnarray}\label{eq:defomega}
&&\Omega = \omega^{(\alpha)}\tau^{(\alpha)}; \nonumber\\
&&\tau^{(\alpha)} \in \left\{\Gamma^{(\beta)}, \Gamma^{(\beta)}T^a\right\};\nonumber\\
&&\Gamma^{(\beta)} \in \left\{1, i\gamma^5, \gamma^\mu,
i\gamma^\mu\gamma^5\right\}.
\end{eqnarray}
The antisymmetric tensor section in $\Omega$ decouples from the theory because of the identity 
$\gamma_\mu\widehat{u}\widehat{v}\gamma^{\mu} = 4u_\mu{}v^{\mu}$

The Bogolyubov self-consistent field approach implies that we should study the
lagrangian 
\begin{equation}
\tilde{\mathcal{L}} = \bar{\psi}\left(i\widehat{\partial} -
\gamma_{\mu}T^a\Omega_c\gamma^{\mu}T^a\right)\psi. \label{eq:meanfield}
\end{equation}
where $\Omega_c$ is a solution of the classical equations of motion for the effective lagrangian obtained by integrating the quarks out of the action (\ref{eq:modaction}).
The integration yields
\begin{equation}
Z = 
C\int\mathcal{D}\Omega\frac{\exp\left\{  iG^2\int{}d^4{x}Sp\gamma_{\mu}T^a\Omega\gamma^{\mu}T^a\Omega\right\}}{Det\left[i\widehat{\partial} - \gamma_{\mu}T^a\Omega\gamma^{\mu}T^a\right]}, 
\end{equation}
which implies
\begin{equation}
S_{eff} = iTr\ln\left[i\widehat{\partial} -
\gamma_{\mu}T^a\Omega\gamma^{\mu}T^a\right] - G^2\int{}d^4{x}Sp\gamma_{\mu}T^a\Omega\gamma^{\mu}T^a\Omega.
\label{eq:effaction}
\end{equation}

The corresponding equation of motion (the gap equation) is
\begin{equation}
\left\langle{x}\left|\gamma_{\mu}T^a\frac{i}{i\widehat{\partial} -
\gamma_{\mu}T^a\Omega\gamma^{\mu}T^a}\gamma^{\mu}T^a\right|x\right\rangle - 2G^2\gamma_{\mu}T^a\Omega(x)\gamma^{\mu}T^a = 0.
\label{eq:gap}
\end{equation}

If we assume $\Omega$ to be a constant scalar, then it's just the case studied in \cite{NJL}, 
which corresponds to quark mass generation $\gamma_{\mu}T^a\Omega\gamma^{\mu}T^a = M$. 
But our goal is to study the inhomogeneous case. 
So we are going to put $\gamma_{\mu}T^a\Omega\gamma^{\mu}T^a = M - \Bomega$ 
and expand the logarithm in (\ref{eq:effaction}) around the constant solution. 
We obtain, omitting the constant part $iTr\ln{i\left(i\hat{\partial} - M\right)}$,
\begin{eqnarray}S_{eff}\left[\Omega\right] &=&  
iTr\ln\left(1 + \widehat{S}\circ\Bomega\right) - 
G^2\int\,d^4x{}\Phi_2\left(\Omega\right) = \nonumber\\&=& 
- G^2\int\,d^4x{}\Phi_2\left(\Omega\right) + 
iSp\widehat{S}(0)\int\,d^4x\Bomega(x) -\nonumber\\ 
&-&\frac{i}{2}Sp\int\,d^4xd^4y\widehat{S}(x-y)\Bomega(y)\widehat{S}(y-x)\Bomega(x)+...
\label{eq:series}
\end{eqnarray}
where $\widehat{S}(x-y)$ is the plain Dirac propagator $\left(i\widehat{\partial}_x - M\right)\widehat{S}(x-y) = \delta(x-y)$ 
and we have defined

\begin{equation}
Sp\gamma_{\mu}T^a\Omega(x)\gamma^{\mu}T^a\Omega(x) \equiv
\Phi_2\left(\Omega\right). \end{equation}.

The first term in the expansion can be rewritten as follows:
\begin{equation}
Sp\widehat{S}(0)\int\,d^4x\Bomega(x) = 
Sp\int\frac{d^4xd^4p}{(2\pi)^4}\frac{\hat{p}+M}{p^2-M^2}\Bomega(x).
\end{equation}
Let us now demonstrate that each term in the series can be reduced to 
\begin{equation}\int\,d^4pd^4x\sum\limits_{i =
0}^{\infty}\widehat{\Xi}_i(x,p).$$ For this, it's enough to demonstrate that
$$\int\,d^4y\widehat{S}(x-y)\Bomega(y)\int\,d^4pe^{-ip(y-z)}\widehat{\Xi}(p,z) = 
\int\,d^4pe^{-ip(x-z)}\widehat{\widetilde{\Xi}}(p,z).\end{equation}
So
\begin{eqnarray}
&&\int\,d^4y\widehat{S}(x-y)\Bomega(y)\int\,d^4pe^{-ip(y-z)}\widehat{\Xi}(p,z) =\nonumber\\&&= \int\frac{d^4yd^4qd^4p}{(2\pi)^4}e^{-iq(x-y)}\frac{\hat{q}+M}{{q}^2-M^2}\Bomega(y)e^{-ip(y-z)}\widehat{\Xi}(p,z) = \nonumber\\
&&=\int\frac{d^4yd^4qd^4p}{(2\pi)^4}e^{-iq(x-y)}\frac{\hat{q}+M}{{q}^2-M^2}\times\nonumber\\&&\times
\sum\limits^{\infty}_{n=0}\frac{(y-z)^{\alpha_1}..(y-z)^{\alpha_n}}{n!}\left[\frac{\partial}{\partial{z}^{\alpha_1}}..\frac{\partial}{\partial{z}^{\alpha_n}}\Bomega(z)\right]
e^{-ip(y-z)}\widehat{\Xi}(p,z) = \nonumber\\&&=
\int\frac{d^4yd^4qd^4p}{(2\pi)^4}e^{-iq(x-y)}\frac{\hat{q}+M}{{q}^2-M^2}\times\nonumber\\&&\times\sum\limits^{\infty}_{n=0}\frac{i^n}{n!}\left[\frac{\partial}{\partial{z}^{\alpha_1}}..\frac{\partial}{\partial{z}^{\alpha_n}}\Bomega(z)\right]
\left[\frac{\partial}{\partial{p}_{\alpha_1}}..\frac{\partial}{\partial{p}_{\alpha_n}}e^{-ip(y-z)}\right]\widehat{\Xi}(p,z) = \nonumber\\&&=\int\frac{d^4yd^4qd^4p}{(2\pi)^4}e^{-iq(x-y)}\frac{\hat{q}+M}{{q}^2-M^2}e^{-ip(y-z)}\times\nonumber\\&&\times\sum\limits^{\infty}_{n=0}\frac{(-i)^n}{n!}\left[\frac{\partial}{\partial{z}^{\alpha_1}}..\frac{\partial}{\partial{z}^{\alpha_n}}\Bomega(z)\right]
\frac{\partial}{\partial{p}_{\alpha_1}}..\frac{\partial}{\partial{p}_{\alpha_n}}\widehat{\Xi}(p,z)= \nonumber\\
&&=\int\,d^4qd^4pe^{-i(qx-pz)}\frac{\hat{q}+M}{{q}^2-M^2}\int\frac{d^4y}{(2\pi)^4}e^{-i(p-q)y}\times\nonumber\\&&\times\sum\limits^{\infty}_{n=0}\frac{(-i)^n}{n!}\left[\frac{\partial}{\partial{z}^{\alpha_1}}..\frac{\partial}{\partial{z}^{\alpha_n}}\Bomega(z)\right]
\frac{\partial}{\partial{p}_{\alpha_1}}..\frac{\partial}{\partial{p}_{\alpha_n}}\widehat{\Xi}(p,z)=\nonumber\\&&=\int\,d^4qd^4pe^{-i(qx-pz)}\frac{\hat{q}+M}{{q}^2-M^2}\delta(p-q)\sum\limits^{\infty}_{n=0}\frac{(-i)^n}{n!}\left[\frac{\partial}{\partial{z}^{\alpha_1}}..\frac{\partial}{\partial{z}^{\alpha_n}}\Bomega(z)\right]
\frac{\partial}{\partial{p}_{\alpha_1}}..\frac{\partial}{\partial{p}_{\alpha_n}}\widehat{\Xi}(p,z)=\nonumber\\&&=\int\,d^4pe^{-ip(x-z)}\frac{\hat{p}+M}{{p}^2-M^2}\sum\limits^{\infty}_{n=0}\frac{(-i)^n}{n!}\left[\frac{\partial}{\partial{z}^{\alpha_1}}..\frac{\partial}{\partial{z}^{\alpha_n}}\Bomega(z)\right]\frac{\partial}{\partial{p}_{\alpha_1}}..\frac{\partial}{\partial{p}_{\alpha_n}}\widehat{\Xi}(p,z),\hspace{0.7cm}Q.E.D.
\end{eqnarray}

Therefore, it's now obvious that if we define the operator 
\begin{equation}
 \widehat{V}(x,p) = \sum\limits^{\infty}_{n=0}\frac{(-i)^n}{n!}\left[\frac{\partial}{\partial{z}^{\alpha_1}}..\frac{\partial}{\partial{z}^{\alpha_n}}\Bomega(z)\right]
\frac{\partial}{\partial{p}_{\alpha_1}}..\frac{\partial}{\partial{p}_{\alpha_n}},
\label{eq:vertex}
\end{equation}
it's possible to express the expansion (\ref{eq:series}) as
\begin{equation}
 S_{eff}\left[\Omega\right] = - G^2\int\,d^4x \Phi_2\left(\Omega\right) + 
i\int\frac{d^4xd^4p}{(2\pi)^4}Sp\sum\limits_{k=1}^\infty\frac{(-1)^{k -
1}}{k}\left[\frac{\widehat{p} + M}{p^2 - M^2}\widehat{V}(x,p)\right]^k\circ1.
\label{eq:loop}
\end{equation}

We can easily read from this that all the integrations in (\ref{eq:series}) reduce to expressions of type
\begin{equation}
I(m,n)_{\alpha_1..\alpha_m} = 
\int\,d^4p\frac{p_{\alpha_1}..p_{\alpha_m}}{\left(p^2 - M^2\right)^n}.
\label{eq:defint}
\end{equation}

First of all, it's easy to see that integrals like this are nonzero for even values of $m$ only and are finite if
\begin{equation}d(n,m) = 2n - m - 4 > 0.\end{equation}
It's obvious that differentiation $\frac{\partial}{\partial{}p^{\mu}}$ of the integrand in (\ref{eq:defint}) 
increments its $d(n,m)$ by $1$ while leaving the difference $n-m$ unchanged. 
Thus we can conclude from (\ref{eq:loop}) that the series contains only the integrals (\ref{eq:defint}) with $m \leq n$. 
But it's also true that a $k$-th order in $\Bomega$ term can contain only the integrals with $n \geq k$.

Therefore we immediately conclude that the expansion (\ref{eq:loop}) has only finite number of divergent terms which are
\begin{enumerate}
\item first order term;
\item second order up to second derivatives of $\Bomega$;
\item third order up to first derivatives;
\item fourth order with no derivatives of $\Bomega$.
\end{enumerate}

Regularization by cutoff will replace the divergencies with some finite factors of order $\Lambda^n$ in the cutoff parameter. Therefore, the finite part can be treated as a small correction to the divergent one. 
So by omitting the finite terms we can build an approximation of the effective
action (\ref{eq:effaction}) that has remarkable features. First, it's similar in
structure to the Ginzburg-Landau functional in superconductivity, since it contains derivatives up to second and nonlinearities up to fourth order; second, it requires no extra conditions being imposed on the model to be valid.

To calculate the terms of our interest we use the Passarino-Veltman reduction \cite{PasVel}
\begin{eqnarray}
&&I(2m+1,n)_{\alpha_1..\alpha_{2m + 1}} = 0;\nonumber\\
&&I(2m,n)_{\alpha_1..\alpha_{2m}} =
CS\left(\eta_{\alpha_1\alpha_2}..\eta_{\alpha_{2m-1}\alpha_{2m}}\right);
\int{}d^4p\frac{p^{2m}}{\left[p^2 - M^2\right]^n}.
\end{eqnarray}
where $C$ is a symmetry factor and $S\left(\eta_{\alpha_1\alpha_2}..\eta_{\alpha_{2m-1}\alpha_{2m}}\right)$ 
is a symmetric tensor power of the Lorentz metrics.

After rather long computation we obtain the following expression for the approximate "Ginzburg-Landau" action:
\begin{eqnarray}
&&\Xi = - G^2\int{d^4x}\Phi_2\left(\Omega\right) + \nonumber\\
&&+ \frac{\ln{\left(\frac{\Lambda^2}{M^2} + 1\right)}}{32\pi^2}\int{}d^4xSp\left\{2M\left[Z(\Lambda) - M^2\right]\Bomega\left(x\right) + M^2\left[\Bomega(x)\right]^2 -\right.\nonumber\\
&&\left. - \frac{Z(\Lambda) - 2M^2}{4}\gamma_\mu\Bomega(x)\gamma^\mu\Bomega(x)
 + M\Bomega(x)i\widehat{\partial}\Bomega(x) - \right. \nonumber\\
&&\left.- \frac{1}{6}\left[\widehat{\partial}\Bomega(x)\right]^2 - \frac{1}{12}\gamma_\mu\left(\partial_\nu\Bomega(x)\right)\gamma^\mu\partial^\nu\Bomega(x) - \right.\nonumber\\
&&\left.- \frac{M}{2}\Bomega(x)\gamma_\mu\Bomega(x)\gamma^\mu\Bomega(x) + \right.\nonumber\\
&&\left. + \frac{i}{6}\gamma_\mu\Bomega(x)\gamma^\mu\left[\partial_\nu\Bomega(x)\right]\gamma^\nu\Bomega(x) -
 \frac{i}{6}\gamma_\mu\Bomega(x)\gamma^\mu\Bomega(x)\widehat{\partial}\Bomega(x) + \right.\nonumber\\
&&\left. + \frac{1}{24}Sp\left[\gamma_\mu\Bomega(x)\gamma^\mu\Bomega(x)\right]^2 
 + \frac{1}{48}Sp\gamma_\mu\Bomega(x)\gamma_\nu\Bomega(x)\gamma^\mu\Bomega(x)\gamma^\nu\Bomega(x)\right\},
\label{eq:action}
\end{eqnarray}
where it has been put
\begin{equation}
 Z(\Lambda) = \frac{\Lambda^2}{\ln\left(\frac{\Lambda^2}{M^2} + 1\right)}.
\end{equation}
Note that differential and nonlinear terms are of the same logarithmic order in $\Lambda$. This implies that nonlinearities play an important role in the inhomogeneous case and can't be treated as just small corrections.

Let us now change variables back by substituting $\Bomega \rightarrow M - \Bomega$. In the fourth order we get

\begin{eqnarray}
&&Sp\gamma_\mu\Bomega\gamma_\nu\Bomega
\gamma^\mu\Bomega\gamma^\nu\Bomega \rightarrow 
Sp\left[\gamma_\mu\Bomega\gamma_\nu\Bomega
\gamma^\mu\Bomega\gamma^\nu\Bomega - \right.\nonumber\\  && \left.- 8M\Bomega\gamma_\nu\Bomega\gamma^{\nu}\Bomega + 4M\gamma_\mu\Bomega\gamma^\mu\Bomega\gamma_\nu\Bomega\gamma^{\nu}
- 8M^2\gamma_\mu\Bomega\gamma^\mu\Bomega +\right.\nonumber\\
&&\left.+ 16M^2\Bomega^2 - 2M^2\left(\gamma_\mu\Bomega\gamma^\mu\right)^2 +
32M^3\Bomega-8M^4\right];\nonumber\\\nonumber\\
&&Sp\left[\gamma_\mu\Bomega(x)\gamma^\mu\Bomega(x)\right]^2 \rightarrow Sp\left[\gamma_\mu\Bomega(x)\gamma^\mu\Bomega(x)\right]^2 + \nonumber\\
&&+ Sp\left[-8M\Bomega\gamma_\mu\Bomega\gamma^\mu\Bomega - 2M\gamma_\mu\Bomega\gamma^\mu\Bomega\gamma_\nu\Bomega\gamma^\nu + \right.\nonumber\\
&&\left.+ 16M^2\Bomega^2 + M^2\left(\gamma_\mu\Bomega\gamma^\mu\right)^2 +
8M^2\gamma_\mu\Bomega\gamma^\mu\Bomega - \right.\nonumber\\ &&\left. - 64M^3\Bomega + 16M^4\right].
\end{eqnarray}

The third order yields

\begin{eqnarray}
&&Sp\Bomega\gamma_\mu\Bomega\gamma^\mu\Bomega \rightarrow -
Sp\left[\Bomega\gamma_\mu\Bomega\gamma^\mu\Bomega -\right.\nonumber\\ &&\left. - 2M\gamma_\mu\Bomega\gamma^\mu\Bomega - 4M\Omega^2 + 12M^2\Bomega - 4M^3\right];\nonumber\\ \nonumber\\
&&Sp\left[\gamma_\mu\Bomega(x)\gamma^\mu\left[\partial_\nu\Bomega(x)\right]\gamma^\nu\Bomega(x) -\gamma_\mu\Bomega(x)\gamma^\mu\Bomega(x)\widehat{\partial}\Bomega(x)\right] \rightarrow \nonumber\\
&&-Sp\left[\gamma_\mu\Bomega(x)\gamma^\mu\left[\partial_\nu\Bomega(x)\right]\gamma^\nu\Bomega(x) -\gamma_\mu\Bomega(x)\gamma^\mu\Bomega(x)\widehat{\partial}\Bomega(x) + \right.\nonumber\\
&&\left. +
M\gamma_\mu\Bomega\gamma^\mu\left[\gamma^\nu,\partial_\nu\Bomega\right] + 4M\Bomega\left[\gamma^\nu,\partial_\nu\Bomega\right] - 4M^2\left[\gamma^\nu,\partial_\nu\Bomega\right]\right].
\end{eqnarray}

Finally

\begin{eqnarray}
&&Sp\Bomega^2 \rightarrow Sp\left[\Bomega^2 - 2M\Bomega + M^2\right];\nonumber\\
&&Sp\gamma_\mu\Bomega\gamma^\mu\Bomega \rightarrow 
Sp\left[\gamma_\mu\Bomega\gamma^\mu\Bomega - 8M\Bomega + 4M^2\right].
\end{eqnarray}

Therefore, after omitting the constants and subtracting full derivatives where appropriate, we obtain

\begin{eqnarray}
&&\Xi = - G^2\int{d^4x}\Phi_2\left(\Omega\right) + \nonumber\\
&&+ \frac{\ln{\left(\frac{\Lambda^2}{M^2} + 1\right)}}{32\pi^2}\int{}d^4xSp\left\{ - \frac{Z(\Lambda)}{4}\gamma_\mu\Bomega(x)\gamma^\mu\Bomega(x)
 - \frac{M}{3}\Bomega(x)i\widehat{\partial}\Bomega(x) - \right. \nonumber\\
&&\left.- \frac{iM}{6}\gamma_\mu\Bomega\gamma^\mu\left[\gamma^\nu,\partial_\nu\Bomega\right] - \frac{1}{6}\left[\widehat{\partial}\Bomega(x)\right]^2 - \frac{1}{12}\gamma_\mu\left(\partial_\nu\Bomega(x)\right)\gamma^\mu\partial^\nu\Bomega(x) - \right.\nonumber\\
&&\left. - \frac{i}{6}\gamma_\mu\Bomega(x)\gamma^\mu\left[\partial_\nu\Bomega(x)\right]\gamma^\nu\Bomega(x) + \frac{i}{6}\gamma_\mu\Bomega(x)\gamma^\mu\Bomega(x)\widehat{\partial}\Bomega(x) + \right.\nonumber\\
&&\left. + \frac{1}{24}Sp\left[\gamma_\mu\Bomega(x)\gamma^\mu\Bomega(x)\right]^2 
 +
 \frac{1}{48}Sp\gamma_\mu\Bomega(x)\gamma_\nu\Bomega(x)\gamma^\mu\Bomega(x)\gamma^\nu\Bomega(x)\right\}.
\end{eqnarray}

Now we note that 
\begin{eqnarray}
&&\Bomega = \phi + \gamma^\mu{}W_\mu; \nonumber\\
&&\phi,W_\mu \in \left\{f + F^a{}T^a + i\gamma^5\left(g + G^aT^a\right)\right|
\left.f,F^a,g,G^a \in \mathbb{R}\right\}.
\label{eq:defalgebra}
\end{eqnarray}
Then it immediately follows from the $\gamma$-matrices' properties that
\begin{eqnarray}
&&\gamma^\mu\phi = \phi^+\gamma^\mu; \nonumber\\
&&\gamma^\mu{}W_\mu = W_\mu^+\gamma^\mu; \nonumber\\
&&\gamma_\mu\Bomega\gamma^\mu = 4\phi^{+} - 2\gamma^\mu{}W_\mu^+.
\end{eqnarray}
Using the identities above we can easily demonstrate that 
\begin{equation}\int{}d^4xSp\left[\frac{M}{3}\Bomega(x)i\widehat{\partial}\Bomega(x)
-
\frac{iM}{6}\gamma_\mu\Bomega\gamma^\mu\left[\gamma^\nu,\partial_\nu\Bomega\right]\right]
\equiv 0\end{equation}
 and the action becomes
\begin{eqnarray}
\label{eq:simpleaction}
&&\Xi = - G^2\int{d^4x}\Phi_2\left(\Omega\right) + \nonumber\\
&&+ \frac{\ln{\left(\frac{\Lambda^2}{M^2} + 1\right)}}{32\pi^2}\int{}d^4xSp\left\{ - \frac{Z(\Lambda)}{4}\gamma_\mu\Bomega(x)\gamma^\mu\Bomega(x)
 -\right.\nonumber\\
&&\left.- \frac{1}{6}\left[\widehat{\partial}\Bomega(x)\right]^2 - \frac{1}{12}\gamma_\mu\left(\partial_\nu\Bomega(x)\right)\gamma^\mu\partial^\nu\Bomega(x) - \right.\nonumber\\
&&\left. - \frac{i}{6}\gamma_\mu\Bomega(x)\gamma^\mu\left[\partial_\nu\Bomega(x)\right]\gamma^\nu\Bomega(x) + \frac{i}{6}\gamma_\mu\Bomega(x)\gamma^\mu\Bomega(x)\widehat{\partial}\Bomega(x) + \right.\nonumber\\
&&\left. + \frac{1}{24}Sp\left[\gamma_\mu\Bomega(x)\gamma^\mu\Bomega(x)\right]^2 
 +
 \frac{1}{48}Sp\gamma_\mu\Bomega(x)\gamma_\nu\Bomega(x)\gamma^\mu\Bomega(x)\gamma^\nu\Bomega(x)\right\}.
\end{eqnarray}

Despite the simplification, the dynamics defined by (\ref{eq:simpleaction}) is still very complicated 
due to the big number of degrees of freedom. That's why we are going investigate simple special cases. 
First of all, it's obvious that we can drop the color sector putting $\phi
= \xi + i\gamma^5\eta; W_\mu = v_\mu +
i\gamma^5w_\mu$.
The next question is whether it is possible to put $W_\mu = 0$.

The only nonvanishing terms in (\ref{eq:simpleaction}) are those that contain even numbers of $\gamma$-matrices. 
Thus, the only terms linear in $W_\mu$ that would produce a nonvanishing part of the equations on $W_\mu$ 
come from the third-order first-derivative part. However, it's easy to demonstrate that these terms are equal to 
$2iSp\left[\left(W^{\mu+} -
W^\mu\right)\phi^+\partial_\mu\phi\right]$ and the
corresponding part of the motion equation vanishes if $\phi(x) =
\xi(x) \in \mathbb{R}$.

The action for $\xi$ is

\begin{eqnarray}
&&\Xi = -G^2\int{d^4x}\xi^2(x) + \nonumber\\
&&+ \frac{N_c\ln{\left(\frac{\Lambda^2}{M^2} +
1\right)}}{8\pi^2}\int{}d^4x\left\{ - Z(\Lambda)\xi^2(x) -
\frac{1}{2}\partial_\mu\xi\partial^\mu\xi + \frac{1}{2}\xi^4)\right\} = \nonumber\\
&& = - K\int{}d^4x\left\{
\frac{1}{2}\partial_\mu\xi\partial^\mu\xi +\left[Z(\Lambda) +
\frac{8\pi^2G^2}{N_c\ln{\left(\frac{\Lambda^2}{M^2} + 1\right)}}\right]\xi^2 -
\frac{\xi^4}{2}\right\}
\end{eqnarray}
or, if we take into account that M is obtained as a constant solution of the gap equation (\ref{eq:gap}), we can finally see that
\begin{equation}\Xi\left[\xi\right] = K\int{}d^4x\left\{
\frac{1}{2}\partial_\mu\xi\partial^\mu\xi + \left[M^2 +
\frac{16\pi^2G^2}{N_c\ln\left(\frac{\Lambda^2}{M^2} + 1\right)}\right]\xi^2 -
\frac{\xi^4}{2}\right\}.
\label{eq:scalaraction}
\end{equation}

\section{The bound states at a domain wall}
The (\ref{eq:scalaraction}) is nothing other than the real $\phi^4$ model. The
equation of motion is
\begin{equation}
\Box\xi - 2A^2\xi + 2\xi^3 = 0,
\label{eq:phi4}
\end{equation}
where we have denoted $A^2 = M^2 +
\frac{16\pi^2G^2}{N_c\ln\left(\frac{\Lambda^2}{M^2} + 1\right)}$. It has the
well-known kink solution
\begin{equation}
\xi(x) = A\tanh\left[Ax^3\right].
\label{eq:kink}
\end{equation}

This solution interpolates between $\xi = -A$ and $\xi = A$.
These are the two vacuum states of the action (\ref{eq:scalaraction}). Thus, our solution describes a domain wall between regions of space with different vacua.

Let us now solve the Dirac equation with $\phi^4$ kink potential. 
The "wideness" 
parameter of a kink can be absorbed into its "height" by scaling the spacetime
variables appropriately:
\begin{eqnarray}
 \tau &=& Ax^0; \nonumber\\
z &=& Ax^3.\nonumber\\
\label{eq:scaling}
\end{eqnarray}

Thus the equation becomes 
\begin{equation}
 \left[i\gamma^0\partial_{\tau} + i\gamma^3\partial_{z} -
 \mu\tanh\left(z\right)\right]\psi(\tau,z) = 0,
\label{eq:initial}
\end{equation}
where $\mu = 1$ in our case but we are going to keep $\mu$ as an arbitrary parameter for a while.

The next step is putting 
\begin{equation}\psi = e^{-iE\tau}
\left[\begin{array}{c}\boldsymbol{\phi}(z)\\
\boldsymbol{\chi}(z)\end{array}\right].\end{equation}
By choosing the appropriate Dirac matrix representation
we reduce the equation to the following system:
\begin{eqnarray}
 i\sigma_z\frac{d\boldsymbol\chi}{dz} + \left[E -
 \mu\tanh(z)\right]\boldsymbol\phi&=&0;\nonumber\\
i\sigma_z\frac{d\boldsymbol\phi}{dz} + \left[E +
\mu\tanh(z)\right]\boldsymbol\chi&=&0.
\end{eqnarray}
Then we make an ansatz 
\begin{eqnarray}\boldsymbol\phi(z) =
\phi(z)\left|\uparrow\right.\rangle;\nonumber\\
\boldsymbol\chi(z) = \chi(z)\left|\uparrow\right.\rangle,
\label{eq:spinup}
\end{eqnarray}
where
$\sigma_z\left|\uparrow\right.\rangle = \left|\uparrow\right.\rangle$ (we might have chosen
the other eigenvector of the Pauli matrix which wouldn't change the picture much).

The resulting system of equations for $\phi,\chi$ can be written in the following matrix form:
\begin{equation}
 \frac{d}{dz}\left[\begin{array}{c}\phi\\\chi\end{array}\right] =
\left[iE\sigma_x - \mu\tanh(z)\sigma_y\right]\left[\begin{array}{c}\phi\\\chi\end{array}\right]
\label{eq:sigmaeq}
\end{equation}
Now let us define
\begin{eqnarray}
 \left|+\right.\rangle&=&\frac{1}{\sqrt{2}}
\left[\begin{array}{c}1\\i\end{array}\right];\nonumber\\
\left|-\right.\rangle&=&\frac{1}{\sqrt{2}}
\left[\begin{array}{c}1\\-i\end{array}\right].
\end{eqnarray}
These vectors satisfy the identities:
\begin{eqnarray}
 \sigma_y\left|+\right.\rangle&=&\left|+\right.\rangle;\nonumber\\
\sigma_y\left|-\right.\rangle&=&-\left|-\right.\rangle;\nonumber\\
\sigma_x\left|+\right.\rangle&=&i\left|-\right.\rangle;\nonumber\\
\sigma_x\left|-\right.\rangle&=&-i\left|+\right.\rangle.
\end{eqnarray}
We can now make the following ansatz:
\begin{equation}
 \left[\begin{array}{c}\phi\\\chi\end{array}\right] = 
p(z)\cosh^{-\mu}(z)\left|+\right.\rangle +
q(z)\cosh^\mu(z)\left|-\right.\rangle,
\label{eq:ansatz}
\end{equation}
which leads us to

\begin{equation}
 \frac{d}{dz}\left[\begin{array}{c}p(x)\\q(x)\end{array}\right] = 
\left[\begin{array}{cc}0 & E\cosh^{2\mu}(z)\\
 -E\cosh^{-2\mu}(z) & 0\end{array}\right]
\left[\begin{array}{c}p(x)\\q(x)\end{array}\right].
\end{equation}
Then we exclude $q(z)$ from this system of equations and obtain the following:
\begin{eqnarray}
&&q(z) = \frac{\cosh^{-2\mu}(z)}{E}\frac{dp}{dz};\nonumber\\
&&\frac{d^2p}{dz^2} - 2\mu\tanh(z)\frac{dp}{dz} + E^2p(z) = 0. 
\label{eq:excludeq}
\end{eqnarray}
If $E=0$ $p$ and $q$ decouple from each other and we can just put $p = 1, q=0$; 
the other linearly independent solution diverges at the infinities and we don't take it into account.

Finally, we put $\zeta \equiv \sinh(z)$ and the last equation turns into
\begin{equation}
\left(1 + \zeta^2\right)\frac{d^2p}{d\zeta^2} + 
\left(1 - 2\mu\right)\zeta\frac{dp}{d\zeta} + E^2p(\zeta) = 0.
\label{eq:imgegenbauer}
\end{equation}
This equation falls into the hypergeometric class, furthermore, it can be easily proven that there's
a series of polynomial solutions 
that are orthogonal at $(-\infty,\infty)$ with the measure 
\begin{equation}
W(\zeta) = \left(1 + \zeta^2\right)^{-\mu - \frac{1}{2}}.
\label{eq:measure}
\end{equation}
Their spectrum is 
\begin{equation}
 E_n^2 = n\left(2\mu - n\right).
\label{eq:spectrum}
\end{equation}
It can be obtained by studying the asymptotics of the equation and taking into account that the leading order should disappear.
These solutions with $|n| < \mu$ correspond to the bound states.
The Rodrigues formula for these polynomials is 
\begin{equation}
p_n(\zeta) = \left(1 + \zeta^2\right)^{\mu + \frac{1}{2}}\frac{d^n}{d\zeta^n}\left(1 + \zeta^2\right)^{-\mu - \frac{1}{2}}.
\label{eq:defpoly}
\end{equation}
One can note that the polynomials are very similar 
to the well-known 
Gegenbauer polynomials.  
See the appendix for more details.
For the complete analysis of the equation (\ref{eq:imgegenbauer}) in terms of
hypergeometric functions see \cite{Charmchi}.

\section{Soap bubble hadrons or Bogolyubov bag?}

The equation (\ref{eq:phi4}) has two possible vacuum states $\pm{A}$. The kink
solution (\ref{eq:kink}) describes a domain wall between the
different-vacuum regions of space. It depends on one parameter $A$ which defines
both vacuum condensate and thickness of the wall, the latter being proportional
to $A^{-1}$.

Let us now consider a case when a spherical bubble of radius $R$ of$-A$ vacuum
is trapped inside the $+A$-vacuum space region. If the bubble is large enough
(i.e. $R\gg{A^{-1}}$), the domain wall at its edge will behave almost as in the
flat case studied above. This means that there must exist a quark state bound to
the bubble surface.

The stress-energy tensor corresponding to the action (\ref{eq:scalaraction}) is
given by the formula
\nopagebreak
\begin{equation}
T^{\mu\nu} = K\left[\partial^{\mu}\xi\partial^{\nu}\xi
- \frac{\eta^{\mu\nu}}{2}\partial_{\mu}\xi\partial^{\mu}\xi -
\eta^{\mu\nu}A^2\xi^2 + \frac{\eta^{\mu\nu}}{2}\xi^4\right].
\label{eq:stress_energy}
\end{equation}

Energy density is given by its component $T^{00}$. For a stationary
configuration, taking into account the equation of motion we obtain simply

\begin{equation}
\varepsilon = - \frac{K\xi^4}{2}. 
\end{equation}

If we have just a locally-inhomogeneous solution that approaches vacuum value at
infinity, such as our domain wall, the zero energy level must correspond to
vacuum, so the energy will be given by

\begin{equation}
E = \frac{K}{2}\int{d^3x}\left[A^4 - \xi^4(x)\right].
\end{equation}

Substituting the kink solution into it and integrating over the $z$-coordinate
yields

\begin{equation}
E = \int{KA^3}dS,
\label{eq:surface_energy}
\end{equation}
where the integration is carried out over the domain wall surface.

This means that the energy of a large spherical vacuum bubble will be
$4{\pi}KA^3R^2$ and the bubble must be unstable. Such a bubble would collapse
spontaneously.

However, if there are quarks bound to the bubble surface, the picture changes
drastically. When the bubble's radius becomes comparable to $A^{-1}$, its
boundary can no longer be considered thin; the bubble's energy gets distributed
over the whole its volume and the bubble no longer contains $-A$-vacuum.
Instead, it can be approximated by a bubble of $\xi = 0$ inside the $\xi = A$
vacuum.

Such a bubble containing quarks is exactly the Bogolyubov bag (see
\cite{NucleonStruct} for details). The quarks in such a bag have energy
proportional to $R^{-1}$. So, the total energy of the system will have the form
$\alpha R^3 + \frac{\beta}{R}$ and have a minimum at some nonzero value of $R$.

An interesting question arises here: whether a bubble
indeed collapses into a Bogolyubov bag or it is stabilised by quark
pressure at some equilibrium radius.

First of all, it is obvious that to answer this question, we must study a bubble
with $R\gg{A^{-1}}$. Thus, we can treat the boundary as almost flat and solve
the problem perturbatively. 

Let us write the Dirac equation in spherical coordinates following the notation
of \cite{NucleonStruct}). The Dirac spinors will have the form 

\begin{equation}
\psi^\mu_\kappa =
\left[\begin{array}{c}
g(r)\chi^\mu_\kappa\\
f(r)\chi^{-\mu}_{\kappa}
\end{array}\right].
\end{equation}

where $\chi^\mu_\kappa$ are eigenvectors of the operators $K =
\beta\left(\vec{\sigma}\vec{\l} + 1\right)$ and z-component of the full angular
momentum $\vec{j} = \vec{\l} + \frac{\vec{\sigma}}{2}$.

Then the Dirac equation will yield 

\begin{eqnarray}
&&\left[E - M(r)\right]g = i\frac{df}{dr} + i\frac{1-\kappa}{r}f; \nonumber \\
&&\left[E + M(r)\right]f = i\frac{dg}{dr} + i\frac{1 + \kappa}{r}g.
\end{eqnarray}

Now we can define
\begin{equation}
\left[\begin{array}{c}
g(r)\\f(r)
\end{array}\right] \equiv \frac{1}{r}u(r),
\end{equation}
which leads us to
\begin{equation}
Eu = M(r)\sigma_z{u} +i\sigma_x\frac{du}{dr} + \frac{\kappa}{r}\sigma_y{u}.
\label{eq:radialcorrection}
\end{equation}

As $R \gg A^{-1}$, we can approximate $M(r)$ as follows: 
\begin{equation}M(r) =
A\tanh\left[A\left(r - R\right)\right].
\label{eq:approx_potential}
\end{equation}
We are interested in the behaviour near $r = R$. Thus we can
assume $1/r = 1/R$ and treat it as a constant in (\ref{eq:radialcorrection}). 
Making the same ansatz as in (\ref{eq:ansatz}) then leads us to the equation
\begin{equation}
\frac{d^2p}{dz^2} - 2\tanh(z)\frac{dp}{dz} + \left(E^2 -
\frac{\kappa^2}{R^2}\right)p(z) = 0,
\label{eq:energy_correction}
\end{equation}
which differs from (\ref{eq:excludeq}) only by an addition to $E^2$ which means
simply a shift of energy levels. For the zero mode we obtain
\begin{equation}
E = \pm\frac{\kappa}{R}.
\end{equation}

The proper perturbative treatment of the problem, however, requires us to take
into account corrections to (\ref{eq:approx_potential}). We can build the
solution to the radial version of the $\phi^4$ equation (\ref{eq:phi4}) as a
power series in $1/R$ with (\ref{eq:approx_potential}) as the zero order. Obviously, this series will
have a nonzero first order term which must be taken into account to calculate
the exact $1/R$ correction to the zero mode energy.

We will not go into the details of exact computation. It is enough for us to
know that the energy of the bubble with a trapped quark is 
\begin{equation}
E = 4\pi{}KA^3R^2 + \frac{\nu}{R},
\label{eq:bubble_energy}
\end{equation}
where $\nu$ is a constant not depending on $R$. The energy
(\ref{eq:bubble_energy}) will have a minimum at
\begin{equation}
R_{0} = \left(\frac{\nu}{4{\pi}K}\right)^{1/3}A^{-1}.
\end{equation}

This radius depends on the constant $K$ which, in turn, is implicitly related
to the NJL mass gap $M$. Let us study this relation in more detail.

We regularize the equation (\ref{eq:gap}) by an Euclidean cutoff at $p^2 =
\Lambda^2$ to arrive at
\begin{equation}
\Lambda^2 - M^2\ln\left[\frac{\Lambda^2}{M^2} + 1\right] =
\frac{8\pi^2G^2}{N_c}.
\label{eq:gap2}
\end{equation}
As the fourth-order term in (\ref{eq:defmodel}) is proportional to $G^{-2}$, we
will study the small $G$ case $G\ll\Lambda$ which corresponds to strong interaction. 
After expanding the logarithm in (\ref{eq:gap2}) into a power series we see
that in the small $G$ case $\frac{\Lambda}{M}$ must be small as well. Thus, for
an approximate value of $M$ we can leave only the first nonvanishing term
in the expansion which yields
\begin{equation}
M = \frac{\sqrt{N_c}\Lambda^2}{4\pi{}G}.
\end{equation}
This, together with (\ref{eq:scalaraction}), leads us to the conclusion that
with the same degree of accuracy $K = \frac{G^2}{2\Lambda^2}$. This means,
that, if the interaction is strong enough, we can expect the equlibrium configuration
of a system of quarks trapped in a vacuum inhomogeneity to be a "soap bubble"
- in the sense that its energy and particle density is concentrated at the
surface.

\section{Conclusions and discussion}

We have developed a Ginzburg-Landau-like approximation of the NJL gap equation
that allows us to explore inhomogeneous vacuum configurations. We have
demonstarted that this approximate equation of motion has a simple scalar sector
which is nothing other than the $\phi^4$ model. It is well-known that the
$phi^4$ field equation has the kink solution which in our case has the physical meaning
of a domain wall between space regions with opposite quark condensate values.

In such an inhomogeneous vacuum Dirac equation for quark states can be solved
exactly. The general problem 
\begin{equation}
\left[\widehat{\partial} - \mu\tanh(z)\right]\psi = 0
\end{equation} 
has a series of bound states with energies given by
\begin{equation}
 E_n^2 = n\left(2\mu - n\right); |n| < \mu,
\end{equation}
however our case is equivalent to $\mu = 1$ which means that the only bound
state has $E = 0$.

if the domain wall is not infinite and planar but encloses a ball of
opposite-sign vacuum, this bound state energy gets an $O(R^{-1})$ correction
which, in turn, implies that under certain conditions the system can be
stabilised by quark pressure resulting in a peculiar ``soap bubble''
configuration which has its energy and matter concentrated at the surface.

Such ``soap bubble'' states are natural candidate for hadrons. However, such a
model for hadrons fails to explain the hadron mass hierarchy.  

If quark pressure and surface tension are the only forces thet define
bubble size, then the more quarks are trapped in the bubble, the bigger and, as
the domain wall energy is proportional to surface area, heavier it must be.

Thus, the model predicts existence of a single-quark configuration lighter than
a proton which is not observed.

So, modifications to our model are required to explain the real hadrons. One
possibility for such a modification is some residual QCD attraction between
trapped quarks. If such an attraction is strong enough, then, as the quarks are
light, the single-quark configuration will become unstable due to pair
production.

Different bubble surface geometry is another possibility. Strictly speaking,
sphericity of the bubble in case of several quarks present has yet to be proven.
It is worth noting here, that if the two-quark bubble is indeed not a sphere but
an elongated ellipsoid stretched along the line that joins the quarks, it will
produce the effective linear potential which is expected for confinement.

\section{Appendix. The fermionic states}

The bound states' wavefunctions should be orthogonal:
\begin{equation}
\int{}d^3x\psi_n^+(x,y,z)\psi_m(x,y,z) = N(n)\delta_{nm}.
\end{equation}

In our case the states are actually "semibound" since the fermion motion 
is restricted along the z-axis alone. Thus the normalization constant will be
infinite.
So we actually should prove that this scalar product reduces to

\begin{equation}
\int{}d^3x\psi_n^+(x,y,z)\psi_m(x,y,z) =\int{dxdy}\int{}dz\psi_n^+(z)\psi_m(z)
\end{equation}
and the z-integral is actually finite and $\int{}dz\psi_n^+(z)\psi_m(z) = C_n\delta_{nm}$ 
for the wavefunctions that don't depend on $x,y$ that were studied above.

To achieve this we should first reconstruct the complete Dirac spinors for the 
(\ref{eq:defpoly}) solutions. Taking into account the transformations from section 3 we obtain
\begin{equation}
 \psi_{n\uparrow}(z) = \frac{\cosh^{-\mu}(z)}{\sqrt{2}}\left[
\begin{array}{c}
\left(p_n(\sinh(z)) +\frac{1}{E_n}\frac{dp_n(\sinh(z))}{dz})\right)|\uparrow\rangle\\
i\left(p_n(\sinh(z)) -\frac{1}{E_n}\frac{dp_n(\sinh(z))}{dz})\right)|\uparrow\rangle
\end{array}\right].
\end{equation}

Thus the normalization condition becomes

\begin{eqnarray}
 &&\langle{n\uparrow}|{m\uparrow}\rangle = \int{dxdy}\int{dz}\cosh^{-2\mu}(z)\times\nonumber\\
&&\times
\left[p_n(\sinh(z))p_m(\sinh(z))
+\frac{1}{E_nE_m}\frac{dp_n(\sinh(z))}{dz}\frac{dp_m(\sinh(z))}{dz}\right].
\end{eqnarray}

However, for the bound states we have, by substituting again $\sinh(z) = \zeta$

\begin{eqnarray}
&&\int{dz}\cosh^{-2\mu}(z)
\left[p_n(\sinh(z))p_m(\sinh(z)) +\frac{1}{E_nE_m}\frac{dp_n(\sinh(z))}{dz}\frac{dp_m(\sinh(z))}{dz}\right]=\nonumber\\
&& = \int\limits_{-\infty}^{+\infty}d\zeta\left(1 + \zeta^2\right)^{-\mu-\frac{1}{2}}
\left[p_n(\zeta)p_m(\zeta) + \frac{1 + \zeta^2}{E_nE_m}\frac{dp_n(\zeta)}{d\zeta}\frac{dp_m(\zeta)}{d\zeta}\right] = \nonumber\\
&&=\int\limits_{-\infty}^{+\infty}d\zeta\left(1 + \zeta^2\right)^{-\mu-\frac{1}{2}}
\left[p_n(\zeta) - 
\frac{1 - 2\mu}{E_nE_m}\zeta\frac{dp_n(\zeta)}{d\zeta} - \frac{1 + \zeta^2}{E_nE_m}\frac{d^2p_n(\zeta)}{d\zeta^2}\right]p_m(\zeta) = \nonumber\\
&& = \left(1 + \frac{E_n}{E_m}\right)\int\limits_{-\infty}^{+\infty}{d\zeta}
\left(1 + \zeta^2\right)^{-\mu-\frac{1}{2}}p_n(\zeta)p_m(\zeta).
\end{eqnarray}

Therefore we can conclude that solutions with $E_n = -E_m$ (we can make a convention 
$E_{-n} =  - E_n, p_{-n}(\zeta) \equiv p_n(\zeta)$) are orthogonal. For the rest of the cases 
scalar product reduces to $\int\limits_{-\infty}^{+\infty}{d\zeta}
\left(1 + \zeta^2\right)^{-\mu-\frac{1}{2}}p_n(\zeta)p_m(\zeta)$ and we can see that the measure is 
exactly (\ref{eq:measure}). 

Only the opposite spin case remains now uninvestigated. However orthogonality of the opposite spin solutions 
is guaranteed trivially by $\langle\uparrow|\downarrow\rangle = 0$ and the spin-down solution are
very similar to the spin-up case studied above, that's why we aren't going into studying them in detail.

\end{document}